\documentclass[aps,prl,showpacs,groupedaddress,amssymb,twocolumn]{revtex4}
\usepackage{bm,amsmath}
\usepackage{times}
\makeatletter
\@ifundefined{pdfoutput} %\% Definitely not using pdftex.
{ %\% Standard TeX
\usepackage[dvips]{graphicx,color}
}
{ %\% Running pdftex.
\ifnum\pdfoutput=0\relax% Are we outputting pdf?
\usepackage[dvips]{graphicx,color}
\fi
\ifnum\pdfoutput=1\relax% Are we outputting pdf?
\usepackage[pdftex]{graphicx,color}
\fi
}
\makeatother
\newcommand{\beq}{\begin{equation}}
\newcommand{\eeq}{\end{equation}}
\newcommand{\beqa}{\begin{eqnarray}}
\newcommand{\eeqa}{\end{eqnarray}}
\newcommand{\nn}{\nonumber \\}

\def \e {\mathrm{e}}

\def \la {\langle}
\def \ra {\rangle}

\def \t {\tau}

\def \Z {{\mathbb Z}}

\def \el {\mathrm{el}}

\def \D {\Delta}

\def \Im {\mathrm{Im} \, }

\begin{document}
%%%%%%%%%%%%%%%%%%%%%%%%%%%%%%%%%%%%%%%%%%%%%%%%%%%%%%%%%%%%%%%
\title{Aharonov--Bohm effect in the non-Abelian quantum Hall fluid}
%%%%%%%%%%%%%%%%%%%%%%%%%%%%%%%%%%%%%%%%%%%%%%%%%%%%%%%%%%%%%%%

\author{Lachezar S. Georgiev$^1$ and Michael R. Geller$^2$}
\affiliation{$^1$Institute for Nuclear Research and Nuclear Energy, 72
Tsarigradsko Chaussee, 1784 Sofia, Bulgaria \\
$^2$Department of Physics and Astronomy, University of Georgia, Athens,
Georgia 30602-2451}

\date{November 8, 2005}

\begin{abstract}
The $\nu=5/2$ fractional quantum Hall effect state has attracted great
interest recently, both as an arena to explore the physics of non-Abelian
quasiparticle excitations, and as a possible architecture for topological
quantum information processing. Here we use the conformal field theoretic
description of the Moore--Read state to provide clear tunneling signatures
of this state in an Aharonov--Bohm geometry. While not probing statistics
directly, the measurements proposed here would provide a first,
experimentally tractable step towards a full characterization of the
5/2 state.
\end{abstract}


\maketitle
\clearpage

%\section{Introduction}
%%%%%%%%%%%%%%%%%%%%%%%%%%

The quantum Hall fluid has become a paradigm of strongly correlated quantum
systems \cite{Prange1990,DasSarma1997}. A combination of two-dimensional
confinement and strong magnetic field leads to rich phenomena driven by
electron-electron interaction and disorder. As such, the use of traditional
theoretical techniques such as many-body perturbation theory has had only
limited success, and, in an approach pioneered in 1983 by Laughlin
\cite{LaughlinPRL83}, some of the most important advances have been made
by correctly guessing the many-particle wave function. The states described
by Laughlin, and their generalizations \cite{Prange1990,DasSarma1997},
have Hall conductances $\sigma_{xy}$ given (in units of $e^2/h$) by
fractions with odd denominators only. The charged excitations, which have
fractional charge \cite{GoldmanSci95,SaminadayarPRL97,dePicciottoNat97}
and statistics \cite{CaminoPRB05}, are abelian anyons. In 1987, however,
evidence for an even-denominator quantized Hall state at $\nu=5/2$ was
discovered in the first excited Landau level \cite{WillettPRL87}, and
the state is now routinely observed in ultrahigh-mobility systems
\cite{PanPRL99,EisensteinPRL02,XiaPRL04}. Motivated in part by this
surprising result, Moore and Read (MR) introduced the ground-state
trial wave function \cite{MooreNPB91}
\begin{equation}
\Psi_{\rm MR}(z_1, z_2, \ldots, z_{N}) = \mathrm{Pf} \left(\frac{1}{z_i-z_j} \right) \prod_{i< j} (z_i-z_j)^2
\label{pfaffian}
\end{equation}
for $N$ (even) electrons in a partially occupied Landau level with
complex coordinates $z_i$, where the first term is the Pfaffian and
the standard Gaussian factor is suppressed. Exact diagonalization
studies \cite{MorfPRL98,RezayiPRL00} indicate that the exact ground
state at $\nu = 5/2$ is close to (\ref{pfaffian}).

Moore and Read also constructed excited-state wave functions. By
identifying (\ref{pfaffian}) with a two-dimensional chiral conformal
field theory (CFT) correlation function
\begin{equation}
\Psi_{\rm MR}(z_1, z_2, \ldots, z_{N})= \langle \psi(z_1) \psi(z_2)  \cdots \psi(z_{N}) \rangle
\label{ground correlator}
\end{equation}
of charge 1 fermion fields
$\psi \equiv \, : \! \! e^{i\sqrt{2}\phi} \! \! : \! \psi_{\scriptscriptstyle {\rm M}},$
where $\phi$ is a $u(1)$ boson \cite{normnote} and
$\psi_{\scriptscriptstyle{\rm M}}$ a neutral Majorana fermion,
MR proposed CFT-based excited states of the form
\begin{equation}
\langle \psi_{\mathrm{qh}}(\eta_1)\cdots  \psi_{\mathrm{qh}}(\eta_{2n}) \psi(z_1) \cdots \psi(z_N) \rangle.
\label{excited correlator}
\end{equation}
Here $\psi_{\rm qh} \equiv : \! \! e^{(i/\sqrt{8})\phi} \! \! \! : \! \sigma$
is the fundamental charge 1/4 quasihole field of the CFT, with $\sigma$
the chiral spin field of the critical Ising model.

The most spectacular prediction of Moore and Read is that the quasiparticle
excitations of the 5/2 state are non-Abelian: An excited state of $2n$
quasiholes has degeneracy $2^{n-1}$, and the braiding of their worldlines
generates elements from the orthogonal group ${\rm SO}(2n)$ acting on the
degenerate multiplet. This property follows from the correlation
function (\ref{excited correlator}), and also from an alternative picture
of the state (\ref{pfaffian}) as a BCS condensate of $l=-1$ pairs of composite
fermions \cite{ReadPRB00}, which supports exotic non-Abelian vortex
excitations \cite{IvanovPRL01,SternPRB04}.

In addition to its intrinsic interest as a system to explore non-Abelian
quantum mechanics, Das Sarma {\it et al.} \cite{DasSarmaPRL05} have proposed
to use a pair of antidots at $\nu=5/2$ to construct a topological quantum
NOT gate, building on an intriguing idea by Kitaev to use the transformations
generated by non-Abelian anyon braiding for fault-tolerant quantum
computation \cite{KitaevAnnPhys03}. Unfortunately, the braiding matrices
generated by (\ref{excited correlator}) are not computationally universal.
But by generalizing the pairing present in $\Psi_{\rm MR}$ to clusters of
$k>2$ particles, Read and Rezayi \cite{ReadPRB99} have proposed a hierarchy
of ground and excited states---CFT correlators of parafermion currents
reducing to (\ref{ground correlator}) and (\ref{excited correlator}) when
$k=2$. These parafermion states are already computationally universal at $k=3$.

Computation with non-Abelian quasiparticles will be incredibly challenging
experimentally. Demonstrating that the actual 5/2 state is in the universality
class of (\ref{pfaffian}), and that the quasiparticles are indeed non-Abelian,
are necessary first steps. Direct interferometric and thermodynamic probes of
the non-Abelian statistics have been proposed recently
\cite{SternPre05,BondersonPre05}. Here we propose an even simpler test of the
MR state, in a similar antidot geometry. Although the tunneling measurement
proposed below does not directly probe the non-Abelian nature of the
excitations, it can distinguish between the tunneling of ordinary electrons
and that of a {\it bosonic} charge 1 excitation we call $\kappa$, which is
allowed by the CFT and which has lower scaling dimension. And the existence
of two inequivalent charge 1 excitations is itself a fingerprint of the
non-Abelian nature of the fundamental quasihole $\psi_{\rm qh}$. Experimental
observation of the electron and $\kappa$ tunneling channels would give
confidence in the MR state, the powerful CFT approach, and, by extension,
the $k>2$ parafermion hierarchy necessary for universal topological quantum
computation.

\begin{figure}[htb]
\centering
\includegraphics*[bb=115 440 490 610,clip,height=4cm]{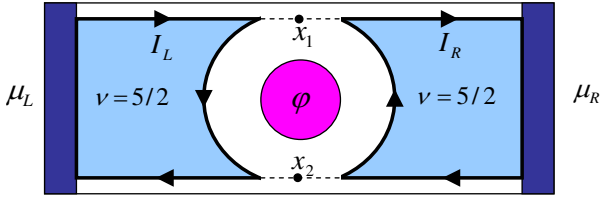}
\caption{(color online). Antidot inside a $\nu = 5/2$ Hall bar with weak
tunneling points at $x_1$ and $x_2$, threaded by an AB flux $\varphi$.}
\label{fig:antidot}
\end{figure}

The geometry we consider is illustrated in Fig.~\ref{fig:antidot}.
The $\nu=1/3$ realization of this system was considered by one of us
previously \cite{Geller&LossPRB97}, and a dual configuration, the double
point-contact interferometer, was studied previously by Chamon
{\it et al.} \cite{ChamonPRB97}. In the strong-antidot-coupling regime
pictured, the edge states, indicated by the arrows, are strongly reflected
by the antidot. This regime can be realized experimentally in two physically
distinct ways: (i) The Hall fluid can be pinched off near the $x_i$, leaving
large tunneling barriers or ``vacuum" regions. Only electrons can tunnel
through these vacuum regions. (ii) The system can begin in the
weak-antidot-coupling regime, where current from the upper edge tunnels
through the antidot (acting as a macroscopic impurity) to the lower one,
and the temperature is then lowered. The weak tunneling regime, which
permits quasiparticle tunneling, is unstable at low temperatures, as in
the $\nu=1/3$ quantum point contact \cite{MoonPRL93}, and the system then
flows to the stable strong-antidot-coupling fixed point, which can be
described by an {\it effective} weak-tunneling theory \cite{fixedpointnote}.
In the $\nu=1/3$ case, this effective theory contains electron tunneling
only, and is identical to that of case (i). The most dramatic illustration
of this fact comes from the exact Bethe Ansatz solution the chiral Luttinger
liquid model for a $\nu=1/3$ point contact containing only charge 1/3
quasiparticles, which nonetheless describes electron-like tunneling far off
resonance \cite{FendleyPRL95}.

In principle, the effective weak-tunneling theory for case (ii) can be
different than that of (i), and we will consider this possibility here.
Because there is no exact solution available for the $\nu=5/2$ quasiparticle tunneling
model, we will guess the form of the effective theory. Guided by the reasonable condition
that {\it any charge 1 excitation of
the CFT preserving the stability of the fixed point}
can potentially tunnel (the charge requirement introduced to recover the $\nu = 1/3$
result), we need to consider a cluster of at least four fundamental quasiholes
$\psi_{\rm qh}$.
 According to the fusion rule
$\sigma \times \sigma = 1 + \psi_{\scriptscriptstyle{\rm M}}$
\cite{CappelliCMP99,GeorgievNPB03}, the product
$\psi_{\rm qh} \times \psi_{\rm qh}  \times \psi_{\rm qh} \times \psi_{\rm qh} = \psi + \kappa$
yields two distinct tunneling objects, the electron/hole $\psi$, and a
charge 1 boson
\begin{equation}
\kappa \, \equiv \, : \! e^{i \sqrt{2} \phi} \! :
\label{kappa definition}
\end{equation}
whose quantum numbers are summarized in Table \ref{excitation table}.
There are also excitations less relevant than the electron that we do not
need to consider. We emphasize that $\kappa$ is an allowed excitation of
the Pfaffian state satisfying the parity rule
\cite{CappelliCMP99,GeorgievNPB03}. The scaling dimension
of $\kappa$ is 1, and {\it if} it tunnels it will dominate electron
tunneling in the low-temperature limit, leaving a clear experimental
signature.

We turn now to a calculation of the source-drain current
\[
I(V,T,\varphi) = I_0(V,T) + I_{\rm AB}(V,T) \,
\cos\left(\frac{\mu}{\Delta \epsilon} + \varphi\right)
\]
as a function of voltage $V$ and temperature $T$, which  according to our analysis
can be decomposed
into flux-independent and period-one oscillatory Aharonov--Bohm (AB)
components. Here $\mu$ is the mean electrochemical potential of the contacts,
$\Delta \epsilon \equiv 2 \pi v/L$ is the noninteracting level spacing on the
antidot with edge velocity $v$ and circumference $L$, and $\varphi$ is the
AB flux in units of $h/e$.

The Hamiltonian in the strong-antidot-coupling regime is
\begin{equation}  \label{H}
H= H_{\rm L}+ H_{\rm R} + \delta H, \ \ \ {\rm with} \ \ \
\delta H=\sum_{i=1,2} \left(\Gamma_i B_i + \Gamma_i^* B_i^\dagger\right).
\end{equation}
The Hamiltonians for the uncoupled
right- and left-moving edge states of length $L_{\rm sys}$
\[
H_{\rm R}={2 \pi v  \over L_{\rm sys} }\left({L}_0-\frac{c}{24}\right) \quad\mathrm{and} \quad
H_{\rm L}={2 \pi v \over L_{\rm sys} } \left(\bar{L}_0-\frac{c}{24}\right)
\]
  are given by the zero modes of the CFT stress tensors $T(z)$ and $\bar{T}(\bar{z})$,
	\[
	L_0 \equiv \oint\frac{dz}{2\pi i} \, z \, T(z) \quad\mathrm{and} \quad
\bar{L}_0 \equiv \oint\frac{d \bar{z}}{2\pi i} \, \bar{z} \, \bar{T}(\bar{z})
\]
 which satisfy the Virasoro algebra with central charge $c=3/2$
\cite{MilovanovicPRB96,CappelliCMP99,GeorgievNPB03}. Then
\[
L_0 = \frac{1}{2}J_0^2+\sum_{n=1}^\infty J_{-n}J_n +
\sum_{n=1}^\infty (n-\textstyle{\frac{1}{2}}) \
\psi^{\scriptscriptstyle{\rm M}}_{-n+\frac{1}{2}}
\psi^{\scriptscriptstyle{\rm M}}_{n-\frac{1}{2}},
\]
and similarly for $\bar{L}_0.$ The Laurent mode expansion of the Majorana
field with antiperiodic boundary conditions on the cylinder
($z=e^{2\pi i \; {x / L_{\rm sys} }}$) is
\[
\psi_{\scriptscriptstyle{\rm M}}(z) =\sum_{n\in\Z}
\psi^{\scriptscriptstyle{\rm M}}_{n-\frac{1}{2}} z^{-n}
\quad \mathrm{with}
\quad
 \psi^{\scriptscriptstyle{\rm M}}_{n-\frac{1}{2}}=\oint\frac{dz}{2\pi i} \, z^{n-1} \,
 \psi_{\scriptscriptstyle{\rm M}}(z).
\]
The $u(1)$ current $J(z) \equiv i\partial_z \phi $
has mode expansion
\[
J(z)=\sum_{n\in\Z} J_n \, z^{-n-1}, \quad \mathrm{with}
\quad
J_n=\oint\frac{dz}{2\pi i} \, z^n \, J(z).
\]
Here $J_0/\sqrt{2}$ and $\bar{J}_0/\sqrt{2}$ are the usual $g=1/2$ Luttinger liquid number
currents $N_{\rm R}$ and $N_{\rm L}$.
The operators
\[
B_i \equiv \psi_{\rm L}(x_i) \psi_{\rm R}^\dagger(x_i), \quad i=1,2,
\]
entering Eq.~(\ref{H}) are
tunneling operators acting at points $x_i$ in Fig.~\ref{fig:antidot}. The
fields $\psi_{\rm L}$ and $\psi_{\rm R}$ appearing in $B$ depend on the
tunneling object. The tunneling amplitudes at $x_1$ and $x_2$ are
\[
\Gamma_1 = \Gamma\,  \e^{i\pi\left(\mu/\D \epsilon +\varphi\right)}
\quad \mathrm{and}\quad
\Gamma_2=\Gamma \, \e^{-i\pi\left(\mu/\D \epsilon +\varphi\right)},
\]
where, with no loss of generality, $\Gamma$ can be taken to be real. The
bare amplitude $\Gamma$ depends on microscopic details and type of tunneling
object.

\begin{table}
\caption{\label{excitation table}
Some chiral conformal fields and their quantum numbers. $\Delta_{\rm c}$
and $\Delta_0$ are the scaling dimensions in the charged and neutral sectors,
$\Delta \equiv \Delta_{\rm c}+\Delta_0$ is the total CFT dimension, and
$\theta \equiv 2 \pi \Delta \ ({\rm mod} \, 2 \pi)$ is the statistics angle.}
\begin{ruledtabular}
\begin{tabular}{|c|c|ccc|c|}
field & charge & $\Delta_{\rm c}$ & $\Delta_0$ & $\Delta$ & $\theta / \pi$ \\ \hline
quasihole $\psi_{\rm qh}$ & $\frac{1}{4}$ & $\frac{1}{16}$ & $\frac{1}{16}$ & $\frac{1}{8}$ & $\frac{1}{4}$ \\
$\kappa$ particle & 1 & 1& 0 & 1& 0 \\
Majorana $\psi_{\scriptscriptstyle{\rm M}}$ & 0 & 0& $\frac{1}{2}$ & $\frac{1}{2}$& 1 \\
electron $\psi$ & 1 & 1& $\frac{1}{2}$ &$\frac{3}{2}$ & 1 \\
\end{tabular}
\end{ruledtabular}
\end{table}

The tunneling current $I(V,T,\varphi)$ can be calculated by linear response theory
along the lines of Refs.~[\onlinecite{Geller&LossPRB97}] and
[\onlinecite{ChamonPRB97}], leading to
\beqa
I_0(V,T) &=& 4 e \Gamma^2 \Im X_{11}(\omega=eV) \quad \mathrm{and}   \nn
I_{\rm AB}(V,T) &=& 4 e \Gamma^2 \Im X_{12}(\omega=eV) , \nonumber
\eeqa
where $X_{ij}(\omega)$ is the
Fourier transform of the response function
\[
X_{ij}(t) \equiv -i \theta(t) \langle [ B_i(t), B_j^\dagger(0) ] \rangle_\beta   ,
\]
with $B_i(t) \equiv e^{i H_0 t}\,  B_i \, e^{-i H_0 t}$
and the thermal average
\[
\la A \ra_\beta =\frac{\mathrm{tr} \ A \, e^{-\beta H_0}}{\mathrm{tr} \ e^{-\beta H_0} }
\]
is taken with respect to the Hamiltonian
\[
H_0 \equiv H_{\rm R} + H_{\rm L}  - \mu_{\rm R} N_{\rm R}  - \mu_{\rm L} N_{\rm L}.
\]
The finite-temperature correlation function $X_{ij}(t)$ is computed in
three steps: (i) First, it is split into products of finite-temperature
correlation functions (with chirality $\pm$) of the form
$\langle \psi_{\pm}^\dagger(x,t)\psi_{\pm}(x',t') \rangle_\beta;$ (ii) Then
these one-dimensional thermal correlation functions are obtained as
zero-temperature correlation functions (after Wick rotation to imaginary
time $\tau$) on a cylinder with circumference
$L_{\rm T} \equiv v/(k_{\rm B} T)$;
(iii) Finally, we map the cylinder to the complex plane by the conformal
transformation $z_\pm=e^{2\pi( i v \t \pm x )/L_{\rm T}}$ where, for primary
fields with scaling dimension $\Delta$,
\begin{equation}\label{2pt}
\langle \psi_{\pm}^\dagger(z)\psi_{\pm}(z') \rangle = (z-z')^{-2 \Delta} \ \ {\rm for} \ \ |z|>|z'|.
\end{equation}
Under the conformal map a chiral primary field transforms as
$\psi_\pm(x,\tau) \rightarrow (2\pi)^{-1/2} [ i\frac{dz}{d(i v \tau \pm x)}]^{\Delta} \, \psi_\pm(z),$ and by going back to real time we obtain
\begin{equation}
\label{Green}
\langle \psi_{\pm}^\dagger(x,t)\psi_{\pm}(0) \rangle_\beta = \frac{\left(\pm i\pi/L_{\rm T}\right)^{4\Delta}} { 2 \pi \ {\rm sh}^{2\Delta} \left[\pi (x\mp v t \pm i\varepsilon )/L_{\rm T}\right]},
\end{equation}
where $\varepsilon$ is a positive infinitesimal required by (\ref{2pt}).
Finally, the desired response function is
\begin{widetext}
\begin{equation}
X_{ij}(t)=  -\frac{\theta(t)}{2\pi^2}\left(\frac{\pi}{L_{\rm T}} \right)^{\! \! 4\Delta} {\rm Im} \, \bigg\lbrace
{\rm sh}\left[\pi\left(x_i-x_j+v t+i\varepsilon\right)/L_{\rm T} \right] \ {\rm sh}\left[\pi\left(x_i-x_j-v t-i\varepsilon\right)/L_{\rm T} \right] \bigg\rbrace^{-2 \Delta}.
\end{equation}
\end{widetext}
When $2 \Delta$ is an integer, which will be the case here,  the transform
$X_{ij}(\omega)$ has an infinite number of poles of order $2 \Delta$ and can
be obtained by residue summation. Note that the local response function
which determines the direct current $I_0$ could be obtained as the limit
\[
X_{11}(\omega)=\lim_{|x_1-x_2|\to 0} X_{12}(\omega).
\]

We expect that both the electrons and $\kappa$ particles will contribute to
the observed tunneling current, as will the filled Landau level. If only
electrons tunnel, they will contribute
\begin{widetext}
\begin{equation}
I_0^{(\rm el)} = {e\over h}  {\gamma_{\rm el}^2 \Delta\epsilon \over 240 \pi^2}
\left\{ 64 \left(\frac{T}{T_0}\right)^4 \left( 2\pi \frac{eV}{\Delta\epsilon} \right)+
20  \left(\frac{T}{T_0}\right)^2\left( 2\pi \frac{eV}{\Delta\epsilon} \right)^3  +
 \left( 2\pi \frac{eV}{\Delta\epsilon} \right)^5 \right\},
\label{I0}
\end{equation}
and
\begin{equation}
I_{\rm AB}^{(\rm el)} = \frac{e}{h}  \frac{2 \gamma_{\rm el}^2 \Delta\epsilon}{\pi}
{(T/T_0)^3 \over{{\rm sh}^3(T/T_0)}} \bigg\lbrace
\bigg[ 2\left(\frac{\pi eV}{\Delta\epsilon}\right)^2+ 2 \left(\frac{T}{T_0}\right)^2
\bigg(1-3 {\rm cth}^2\left(\frac{T}{T_0}\right) \bigg) \bigg] \sin\bigg({\pi eV \over \Delta\epsilon}\bigg) +
6 \left(\frac{\pi eV}{\Delta\epsilon}\right) \frac{T}{T_0} {\rm cth}\left(\frac{T}{T_0}\right)
\cos\bigg({\pi eV \over \Delta\epsilon}\bigg) \bigg\rbrace ,
\label{IAB}
\end{equation}
\end{widetext}
which is identical to that of the $g=1/3$ chiral Luttinger liquid
\cite{Geller&LossPRB97,ChamonPRB97}. Here
\begin{equation}
T_0 \equiv \hbar v / \pi k_{\rm B} L,
\label{T0}
\end{equation}
is a temperature scale associated with the level spacing of the antidot
edge state of velocity $v$ and circumference $L$. Above $T_0$, the thermal
length $L_{\rm T}$ becomes smaller than $L$, and the AB oscillations become
washed out. $\gamma \equiv \Gamma/(v L^{2\Delta-1})$ is a dimensionless
tunneling amplitude. Each of the two filled Landau levels also contributes a
  Fermi liquid  (FL) current $I^{(\rm FL)}_0 = (e^2/h) 2\, \gamma_{\rm FL}^2 V$ and
\begin{equation}
I^{(\rm FL)}_{\rm AB} = { e  \over h} 2
\gamma_{\rm FL}^2  {\Delta\epsilon \over \pi}{(T/T_0)   \over {\rm sh}(T/T_0)}
\sin\left(\frac{\pi eV}{\Delta\epsilon}\right).
\label{FL IAB}
\end{equation}

\begin{figure}[htb]
\centering
\caption{(color online) The Aharonov--Bohm current $I_{\rm AB}(V,T)$ for $\Delta=1$
	($\kappa$ tunneling)  as a function of the applied voltage $V$
compared to $\Delta=3/2$  (electron tunneling) reduced by a factor of $100$ and
the chiral Fermi liquid current ($\Delta=1/2$) multiplied by a factor of 10 at $T=T_0$.
\label{fig:I_AB}}
\includegraphics*[bb=20 20 540 450,height=7.5cm]{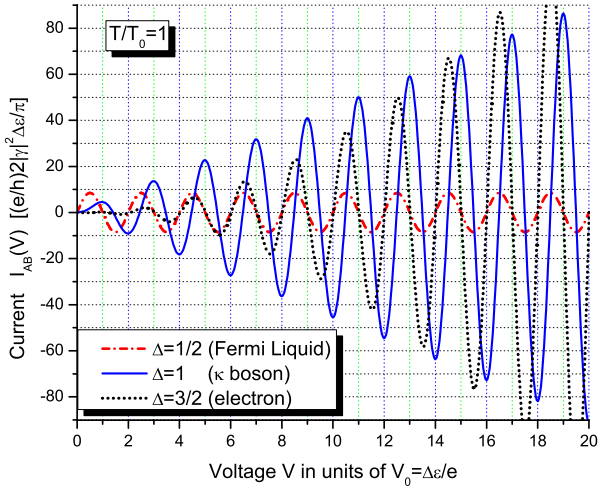}
\end{figure}

The contribution from the $\kappa$ channel alone is
\begin{equation}
I_0^{(\kappa)} = \frac{e^2}{h}\frac{4 \gamma_\kappa^2}{3}  \left(\frac{T}{T_0}\right)^2  V \left[ 1+\frac{1}{4}\left(\frac{eV}{\pi k_{\rm B}T}\right)^2\right],
\end{equation}
and
\begin{widetext}
\begin{equation}
I_{\rm AB}^{(\kappa)} = \frac{e}{h}\frac{2 \gamma_\kappa^2 \D \epsilon}{\pi}
\frac{\left(T/T_0 \right)^2}{{\rm sh}^2\left(T/T_0 \right)}  \left\{  2\left(\frac{T}{T_0}\right){\rm cth}\left(\frac{T}{T_0} \right)
\sin\left(\pi\frac{eV}{\D \epsilon} \right) - 2\pi\frac{eV}{\D \epsilon} \cos\left(\pi\frac{eV}{\D \epsilon} \right) \right\}.
\end{equation}
\end{widetext}
The oscillations of the AB currents as functions of the voltage for the three channels
at temperature $T=T_0$ are shown in Fig.~\ref{fig:I_AB}.

An experiment will observe these two or three transport channels in parallel,
the contribution of each determined by the bare amplitudes $\gamma_\kappa$,
$\gamma_{\rm el},$ and $\gamma_{\rm FL}.$ However, the $\kappa$ channel will
always dominate the electron channel in the low-energy limit.
%%%%%%%%%%%%%%%%%%%%%%%
%%%%%%%%%%%%%%%%%%%%%%

\begin{figure}[htb]
\centering
\includegraphics*[bb=20 20 570 440,height=7cm]{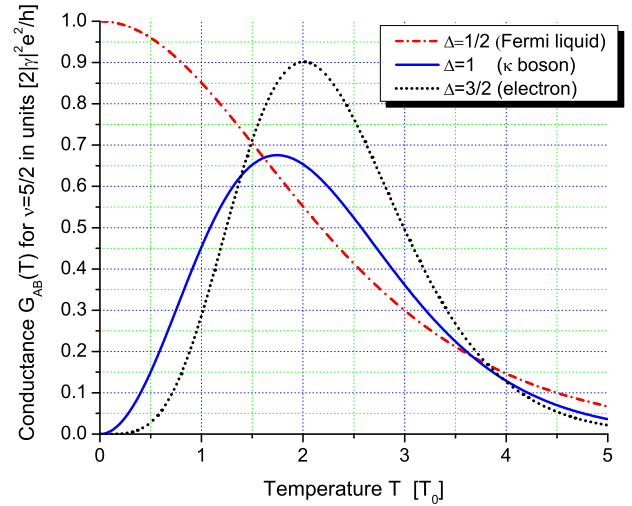}
\caption{(color online). Aharonov--Bohm conductances for electron and
$\kappa$ tunneling at $\nu = 5/2$. The Fermi liquid conductance is also shown.
\label{G_AB}}
\end{figure}

The linear conductance $G \equiv (dI/dV)_{V \rightarrow 0}$ for these channels
takes the form
\[
G(\phi,T)=G_0(T)+\cos\left[2\pi\left({\mu \over \D\epsilon}+\phi\right) \right]G_{\rm AB}(T),
\]
where  $G_0$ and $G_{\rm AB}$ are the direct and the AB conductances, respectively.
For the $\kappa$ channel they read
\beqa
G_0^{(\kappa)}(T)&=&\left(\frac{e^2}{h}\right) \frac{4\gamma_\kappa^2}{3}
 \left(\frac{T}{T_0}\right)^2, \nn
G_{\rm AB}^{(\kappa)}(T)&=&\left(\frac{e^2}{h}\right) 4 \gamma_\kappa^2
\frac{\left(T/T_0\right)^2}{\mathrm{sh}^2\left(T/T_0 \right)}
\left[\left( \frac{T}{T_0}\right) \mathrm{cth}\left( \frac{T}{T_0}\right) -1 \right],
\nonumber
\eeqa
while for the electron channel we obtain
\beqa
G_0^{(\el)}(T)&=&\left(\frac{e^2}{h}\right) 4\gamma_\el^2\left(\frac{4}{15}\right)
 \left(\frac{T}{T_0}\right)^4, \nn
G_{\rm AB}^{(\el)}(T)&=&\left(\frac{e^2}{h}\right) 4 \gamma_\el^2
\frac{\left(T/T_0\right)^3}{\mathrm{sh}^3\left(T/T_0 \right)} \times \nn
&&\left[\left(\frac{T}{T_0}\right)^2 \mathrm{cth}^2\left( \frac{T}{T_0}\right)+
3 \left(\frac{T}{T_0}\right) \mathrm{cth}\left( \frac{T}{T_0}\right)  \right].
\nonumber
\eeqa
 These conductances are compared to the Fermi liquid ones
\beqa
G_0^{({\rm FL})}(T)&=&\left(\frac{e^2}{h}\right) 2\gamma_{\rm FL}^2, \nn
G_{\rm AB}^{({\rm FL})}(T)&=&\left(\frac{e^2}{h}\right) 2 \gamma_{\rm FL}^2
\frac{\left(T/T_0\right)}{\mathrm{sh}\left(T/T_0 \right)}
\nonumber
\eeqa
 and are plotted in Fig.~\ref{G_AB}.
Both the electron and $\kappa$ contributions
display a pronounced maximum as a function of temperature. In the
$T \rightarrow 0$ limit the $\kappa$ contribution, varying as $T^2,$ dominates
the electron contribution. The temperature dependence of $G$ in the low- and
high-temperature regimes is summarized in Table \ref{conductance table}. There
is also an interesting zero-temperature nonlinear regime
$k_{\rm B}T \ll eV \ll k_{\rm B}T_0,$
where both the direct and AB currents vary as $I^{(\kappa)} \sim V^3$ and
$I^{(\rm el)} \sim V^5,$ independent of temperature.
Finally, we also note
that the limit of a single $\nu = 5/2$ quantum point contact (in the stable,
strong tunneling regime) follows from our results by letting
$T_0 \rightarrow \infty$; the electron contribution in this case agrees
with that for tunneling between a FL and $\nu = 5/2$ edge state
\cite{ReadPRB96}.

\begin{table}
\caption{\label{conductance table}
Asymptotic tunneling conductance. $T_0$ is a crossover temperature
defined in Eq.~(\ref{T0}).}
\begin{ruledtabular}
\begin{tabular}{|c|c|c|}
& $eV \ll k_{\rm B}T \ll k_{\rm B}T_0$ & $eV \ll k_{\rm B}T_0 \ll k_{\rm B}T$ \\ \hline
Fermi liquid   & $G_0 \sim {\rm const}$   & $G_0 \sim {\rm const}$  \\
$\Delta = \frac{1}{2}$ & $G_{\rm AB} \sim {\rm const} $  & $G_{\rm AB} \sim T e^{-T/T_0}$ \\ \hline
$\kappa$ particle   & $G_0 \sim T^2$   & $G_0 \sim T^2$  \\
$\Delta = 1$ & $G_{\rm AB} \sim T^2$  & $G_{\rm AB} \sim T^3 e^{-2T/T_0}$ \\ \hline
electron   & $G_0 \sim T^4$   & $G_0 \sim T^4$  \\
$\Delta = \frac{3}{2}$ & $G_{\rm AB} \sim T^4$  & $G_{\rm AB} \sim T^5 e^{-3T/T_0}$ \\
\end{tabular}
\end{ruledtabular}
\end{table}

In conclusion, we have used the CFT picture of Moore and Read
\cite{MooreNPB91} to calculate the AB tunneling spectrum in a $\nu=5/2$
antidot geometry. Observing the electron and possibly $\kappa$ transport
channels will give evidence in support of the non-Abelian nature of the
$5/2$ state.

We thank Ady Stern for useful discussions. LSG has been
partially supported by the FP5-EUCLID Network Program of the EC under
Contract No. HPRN-CT-2002-00325 and by the Bulgarian National Council
for Scientific Research under Contract No.~F-1406. MRG was supported
by the NSF under grants DMR-0093217 and CMS-040403.

\bibliography{/Users/mgeller/Papers/bibliographies/MRGhall,/Users/mgeller/Papers/bibliographies/MRGqc-general,/Users/mgeller/Papers/bibliographies/MRGmanybody,/Users/mgeller/Papers/bibliographies/MRGpre,/Users/mgeller/Papers/bibliographies/MRGbooks,/Users/mgeller/Papers/bibliographies/MRGgroup,abnotes}

\begin{thebibliography}{32}
\bibitem[{nor()}]{normnote}
\bibinfo{note}{This is normalized according to $\langle \phi(z) \phi(z')
  \rangle = -\ln(z-z')$.}
\bibitem[{fix()}]{fixedpointnote}
\bibinfo{note}{This assumes, of course, that there is a stable strong-tunneling
  fixed point here, and that there are no other stable fixed points.}
\end{thebibliography}

\end{document}